\documentclass[aps,prr,amsmath,amssymb,floatfix,twocolumn]{revtex4-1}

\usepackage{multirow}
\usepackage{bbold}
\usepackage{graphicx}
\usepackage{subfigure}
\usepackage{color}
\usepackage{hyperref}
\usepackage[justification=raggedright]{caption}
\usepackage{filecontents}

\begin{document}
\preprint{APS/123-QED}

\title{Identification of soft modes in amorphous Al$_{2}$O$_{3}$ via first-principles}
\author{Alexander C. Tyner$^{1,2}$}
\email{alexander.tyner@su.se}
\author{Joshuah T. Heath$^{1,2}$}
\author{Thue Christian Thann$^{3}$}
\author{Vincent P. Michal$^{3}$}
\author{Peter Krogstrup$^{3}$}
\author{Mark Kamper Svendsen$^{3}$}
\author{Alexander V. Balatsky$^{1,2}$}

\affiliation{$^{1}$ Nordita, KTH Royal Institute of Technology and Stockholm University 106 91 Stockholm, Sweden}
\affiliation{$^2$ Department of Physics, University of Connecticut, Storrs, Connecticut 06269, USA}
\affiliation{$^{3}$NNF Quantum Computing Programme, Niels Bohr Institute, University of Copenhagen, Denmark}

\date{\today}

\begin{abstract} 
\noindent Amorphous Al$_{2}$O$_{3}$ is a fundamental component of modern superconducting qubits. While amorphous oxides offer distinct advantages, such as directional isotropy and a consistent bulk electronic gap, in realistic systems these compounds support {\color{black}two-level systems (TLSs) which} couple to the qubit, expediting decoherence. In this work, we perform a first-principles study of amorphous Al$_{2}$O$_{3}$ {\color{black}and identify low-energy modes in the electronic and vibrational spectra as a possible origin for TLSs.}  
\end{abstract}

\maketitle
\par 
\section{Introduction} 
In recent years, superconducting qubits have emerged as a leading platform for quantum computation in both academia and industry~\cite{kjaergaard2020superconducting,devoret2004superconducting,Transmon,krantz2019quantum}. 
The existence of two-level systems {\color{black}(TLS)} within superconducting qubits have been proposed as a contributing factor to the rate of decoherence~\cite{gordon2014hydrogen,PhysRevB.96.064504,PhysRevB.80.134517,PhysRevB.72.024526,PhysRevA.64.052312,muller2019towards,mansikkamaki2024two,PhysRevX.13.041005,10.1063/5.0017378,Gunnarsson_2013,pritchard2024suppressed,PhysRevB.87.144201}. {\color{black}Despite the presence of TLSs,} amorphous Al$_{2}$O$_{3}$ {\color{black}remains} well suited for use in superconducting qubits, as the lack of long range order has the consequence of making the electronic properties directionally isotropic. {\color{black}Amorphous Al$_{2}$O$_{3}$} can be customized to a variety of architectures without concern for grain-boundaries and other complications of cleaving crystalline compounds\cite{morrissey1984faceted}, and is dynamically stable {\color{black}with robust electronic properties}~\cite{PhysRevLett.103.095501,katiyar2005electrical}.


\par
\begin{figure}
  \centering
    \includegraphics[width=8cm]{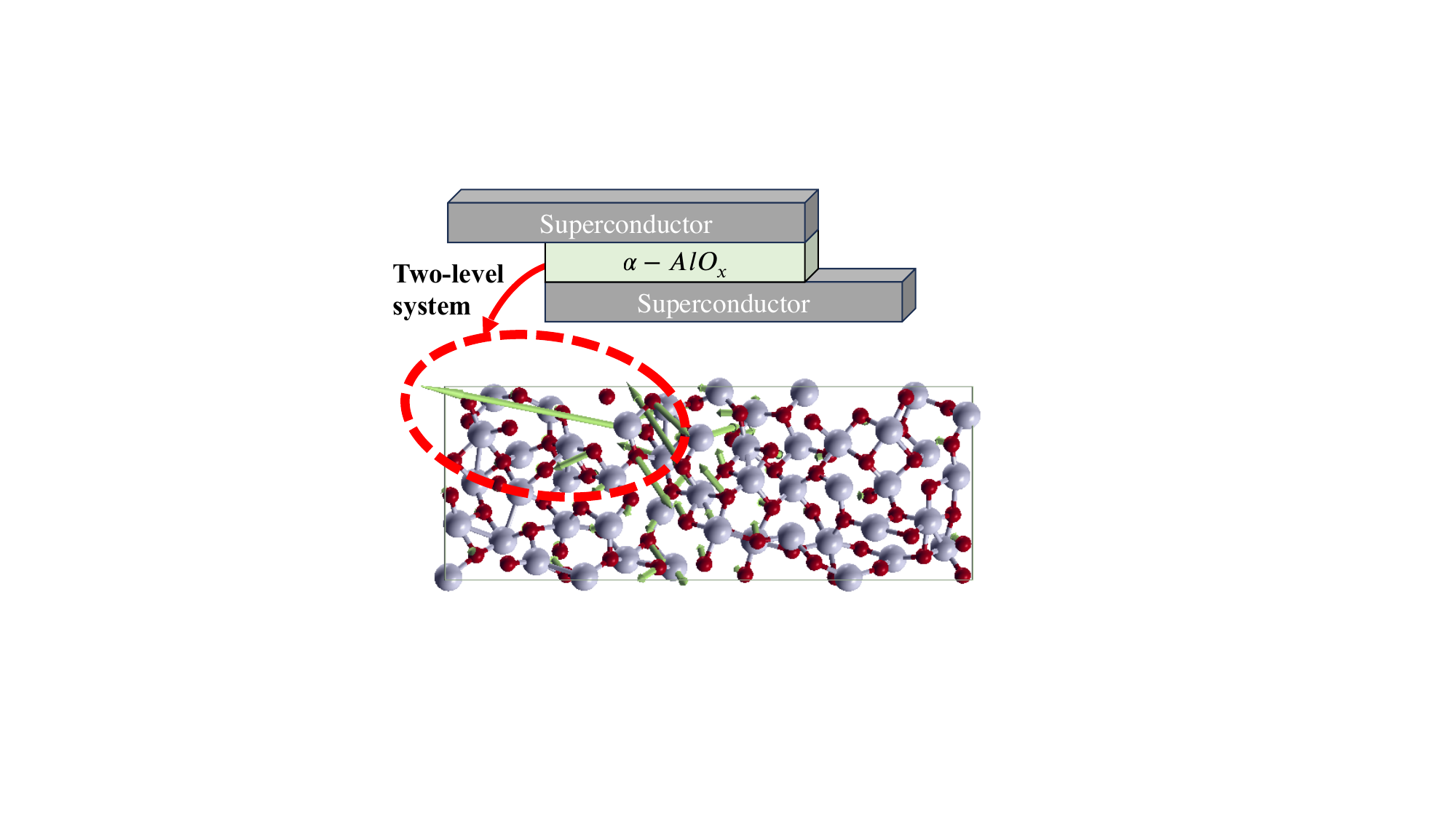}
    \caption{Schematic of a Josephson junction for which amorphous aluminum oxide is used as the insulating barrier. We identify low-frequency modes, vibrons, in real space within the amorphous oxide, corresponding to two-level systems which couple to the qubit and hasten decoherence.  }
    \label{fig:Schem}
\end{figure}

 
The hypothesis we investigate {\color{black} is that localized low-frequency phonons in amorphous Al$_{2}$O$_{3}$, {\color{black}which we will refer to as vibrons and show schematically in Fig. \eqref{fig:Schem}}, may be considered as TLSs}. These ``soft localized modes"  oscillate with  {\color{black}frequencies} that are much lower than {\color{black}the} typical Bose peak and Debye energies. Hence, once quantized, these modes can induce local TLS dynamics and induce decoherence at low frequencies. {\color{black}The hypothesis that spatially localized modes, often described as dangling bonds, can give rise to TLSs has been put forth in other studies\cite{PhysRevB.43.5039,PhysRevLett.66.636,wang2019low,muller2019towards} however, to the best knowledge of the authors, an attempt to directly compute the low-frequency modes of amorphous Al$_{2}$O$_{3}$ via first-principles has yet to be carried out.} {\color{black} To investigate this proposal,}
we generate multiple amorphous samples\cite{cooper2000density,PhysRevB.61.2349,PhysRevB.44.11092,PhysRevLett.60.204}. The starting point for generation of these samples is supercells of the ordered solid. The ordered solids are then melted using Car-Parinello molecular dynamics~\cite{cpmd}. Multiple annealing cycles are then carried out until the samples exhibit stability at room temperature. Prior to computation of electronic properties and phonon modes, relaxation of the atomic positions at zero temperature is performed via density functional theory.

\begin{figure*}
  \centering
    \includegraphics[width=15cm]{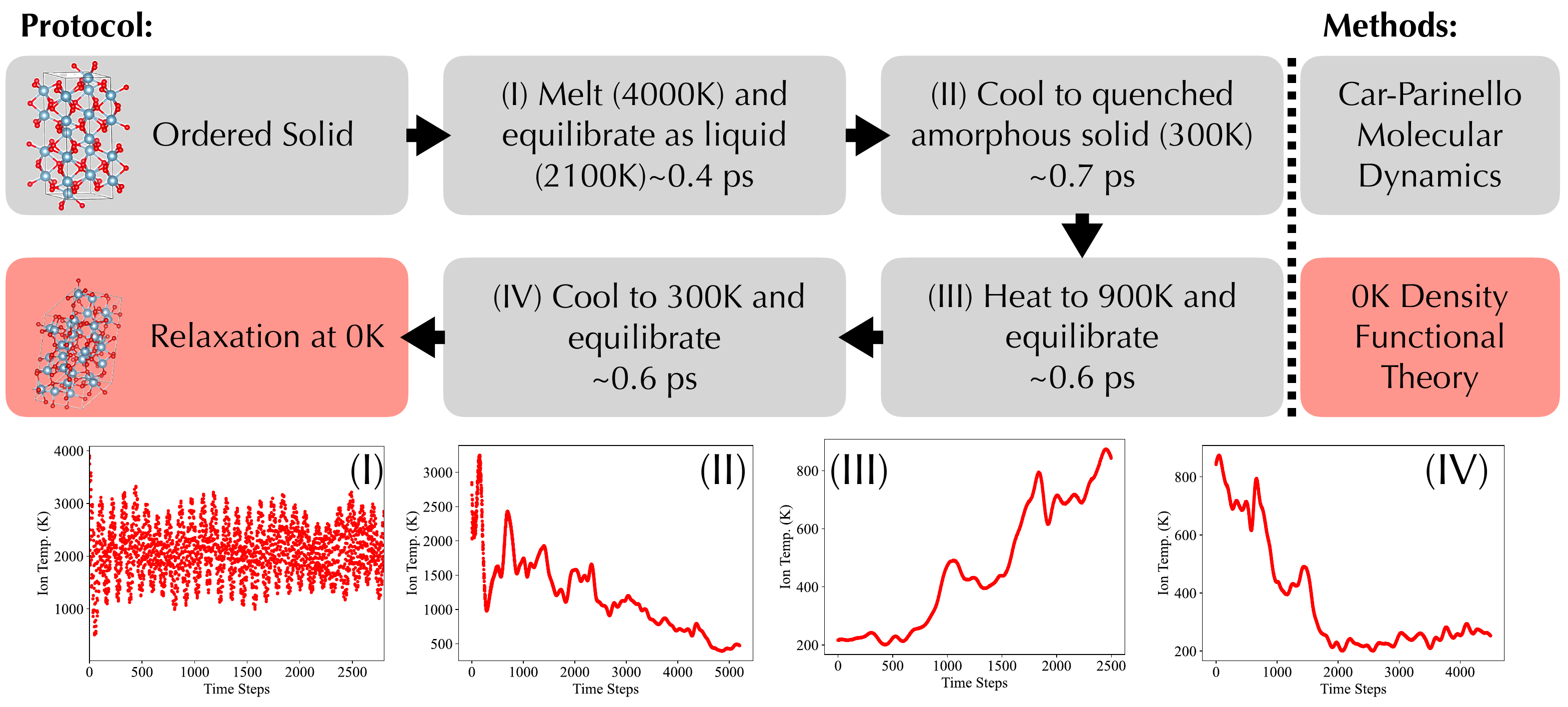}
    \caption{Protocol for generating amorphous samples of Al$_{2}$O$_{3}$ utilizing Car-Parinello molecular dynamics. The samples are then relaxed at zero temperature using density functional theory. }
    \label{fig:protocol}
\end{figure*}
\par 
The low-frequency phonon modes of the resulting amorphous oxides are computed using density functional perturbation theory and mapped in real-space. Our results provide evidence that, when relaxed at zero-temperature, amorphous Al$_{2}$O$_{3}$ support phonon modes at or below 1THz localized in real-space, in alignment with the expectation for a TLS. We chose to examine the window of 0-1 THz rather than the more restricted range of 1-10 GHz at which a qubit generally operates as our first-principles computations encounter finite size effects which can harden soft modes in amorphous solids, shifting frequencies higher\cite{PhysRevB.75.045411}.{\color{black} Nevertheless, we can utilize alternative metrics, namely the dipole-moment, to determine the validity of our hypothesis that these modes act as TLSs. Before concluding, we also discuss potential mitigation strategies through the use of alternate amorphous oxides.}

\section{Generation of amorphous oxides}
\par
Unlike in {\color{black}the} analysis of crystalline systems, structure files are not readily available for amorphous systems. This is because amorphous {\color{black}systems lack long range order, making each sample distinct from all others.} In order to extract universal features we must follow similar procedures to those used to study disordered systems. Namely, we must use as large a system size as possible and average over multiple distinct structures in an attempt to recover universal properties and limit finite size effects. {\color{black}Variations of the finite state samples reflect the variability of the amorphous Al$_{2}$O$_{3}$ structure in realistic junctions.} 
\par 
In order to generate the structures we begin with ordered crystals of Al$_{2}$O$_{3}$ in spacegroup R3$\bar{c}$. A supercell of 120 atoms is formed. 
\begin{figure}
  \centering
    \includegraphics[width=8cm]{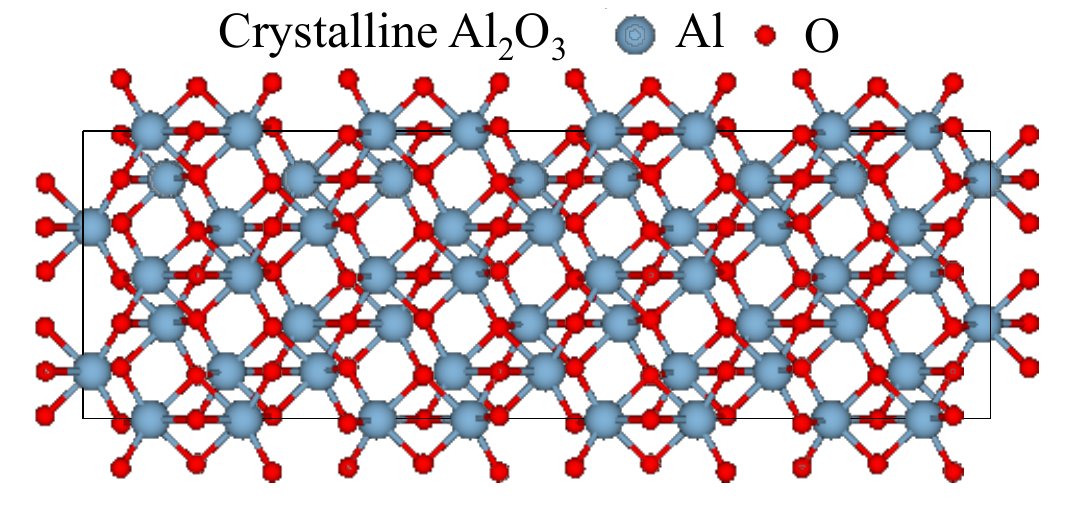}
    \caption{A $2\times 1\times 2$ sample of Al$_{2}$O$_{3}$ in the ordered crystalline phase containing a total of 120 atoms. This is the starting point for the molecular dynamics protocol to generate amorphous samples. }
    \label{fig:SC}
\end{figure}
\begin{figure}
  \centering
    \includegraphics[width=8cm]{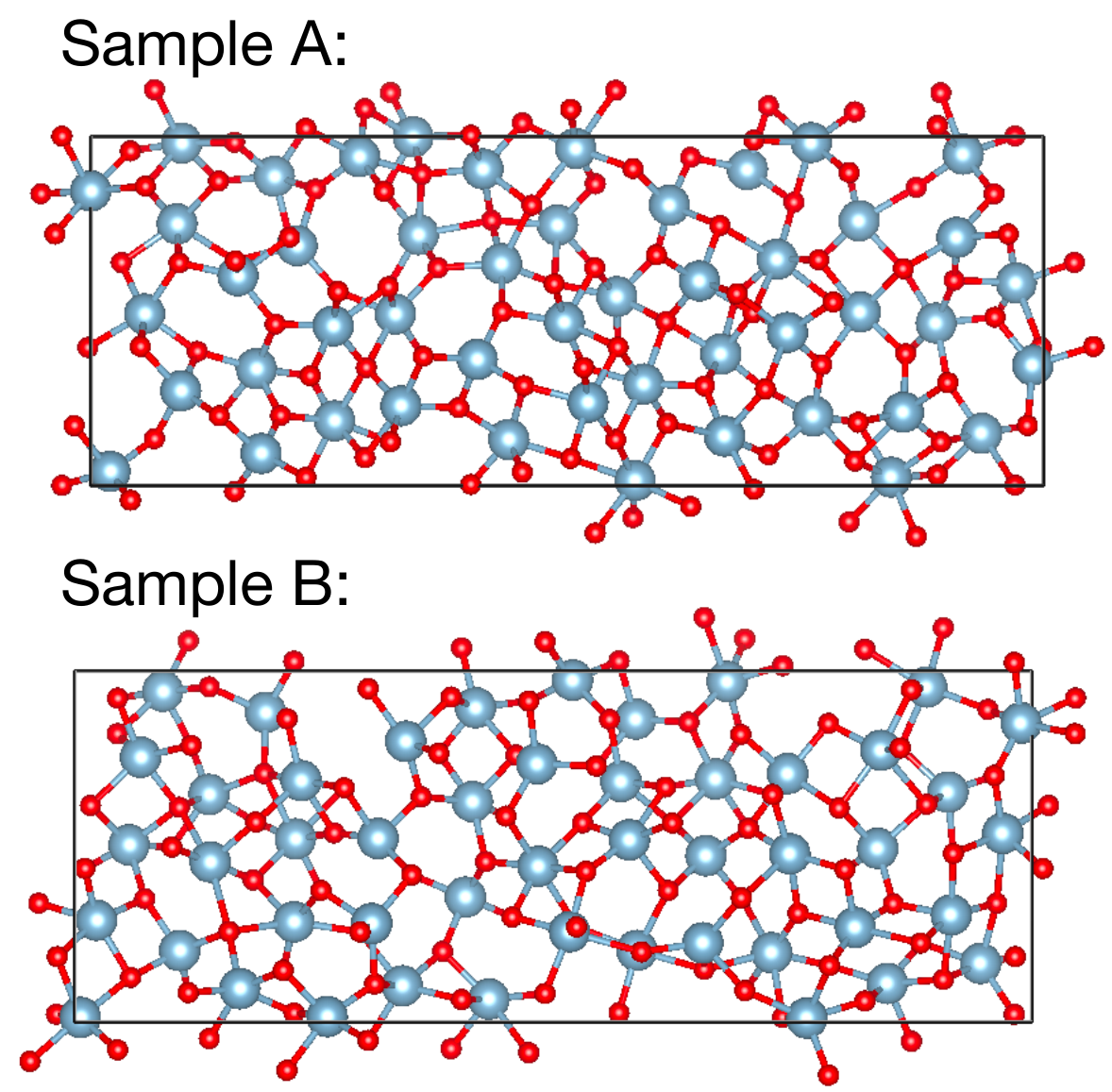}
    \caption{Two of the generated amorphous samples of Al$_{2}$O$_{3}$ at zero temperature. Each sample is unique in accordance with the lack of order in amorphous systems.}
    \label{fig:samples}
\end{figure}
\begin{figure*}
  \centering
    \includegraphics[width=15cm]{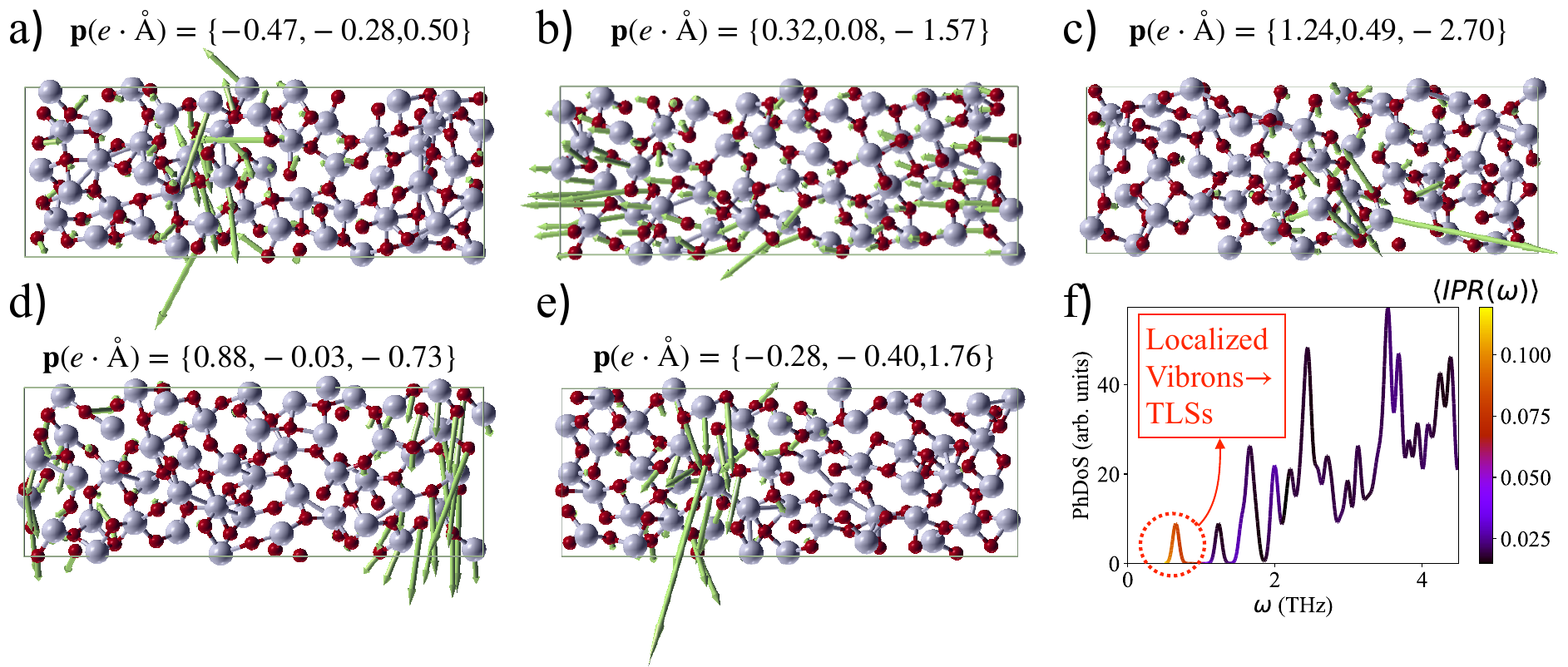}
    \caption{(a)-(e)Displacement vectors are shown for each atom for the lowest frequency phonon modes at the $\Gamma$ location for 5 distinct samples of amorphous Al$_{2}$O$_{3}$. The dipole moment computed for each phonon mode is listed above. The phonon modes all fall within the range 0-1 THz and the modes appear localized in real-space. We label these modes as vibrons. This observation in accordance with magnitude of the electronic dipole moments provide evidence for these phonon modes as two-level systems within a superconducting qubit. (f) Phonon density of states upon averaging over all amorphous samples. {\color{black} The curve is colored according to the value of inverse participation ratio, eq. \eqref{eq:ipr}, averaged over all amorphous samples. Increased IPR indicates enhanced localization. It is clear the modes associated with the low-frequency peak, the vibrons seen in (a)-(e), are significantly more localized than the high-frequency modes. The dipole moment of these modes is in correspondence with the observed dipole moment of TLSs in amorphous aluminum oxide.}}
    \label{fig:LowFModes}
\end{figure*}
\subsection{Generation of amorphous Al$_{2}$O$_{3}$ samples}
\par 
To form the amorphous structures we follow a procedure put forth in Ref. \cite{cooper2000density} for {\color{black}the} generation of amorphous SiO$_{x}$ utilizing Car-Parinello molecular dynamics (CPMD) as implemented within the Quantum Espresso software package\cite{QE-2020,Perdew1996}. The procedure is summarized here and shown schematically in Fig. \eqref{fig:protocol}. First, the atoms are randomly displaced from their starting positions, shown in Fig. \eqref{fig:SC}, by a maximum of 2 percent of the lattice vector magnitude in each direction. Throughout the CPMD computations we use a $5 \times 5 \times 5$ real-space mesh for the wavefunction and charge density FFT. A plane-wave cutoff of 35 Ry, a timestep of 5.5 a.u. (1 a.u.$= 2.4189 \times 10^{-17} s$), a fictitious electron mass of 300 a.u. (1 a.u. = $9.10939 \times 10^{-31}$kg)\cite{cooper2000density} and non-relativistic pseudopotentials from the {\color{black}PseudoDojo}~\cite{van2018pseudodojo} data base are used. 
\par 

After the atomic positions are randomized the structures are melted by rescaling the velocities to reach a temperature of $4000\pm 400 K$, which is above the melting point of $2345 K$\cite{schneider1967effect}, for 200 time steps. In this period the system becomes liquid-like. We then apply Nose-Hoover thermostats to the ions and electrons. The ion thermostat is set to $2100 K$ and a characteristic frequency of 15 THz, determined by the peak in the crystalline phonon density of states. For the electrons the thermostat was set to a target kinetic energy of 0.0075 a.u. and a characteristic frequency of 1350 THz. The liquid phase is equilibrated for 2800 time steps within the canonical ensemble. We then cool this sample at a rate of $2.4 \times 10^{15} K s^{-1}$. At this rate the sample is quenched from $2100 K$ to $300 K$ in $\sim 0.7 ps$. To ensure stability, the amorphous structure is then put through a secondary annealing cycle in which the temperature is raised to $900 K$ over the course of $0.1 ps$, equilibrated at this temperature for $0.53 ps$ and then cooled to $300K$ over $0.25 ps$. Finally, the system is allowed to equilibrate at $300K$ for 3500 time steps. The final structure is then allowed to evolve within an NVE ensemble and does not deviate significantly in temperature after 3500 time steps indicating stability.  
\par
\begin{figure}
  \centering
    \includegraphics[width=9cm]{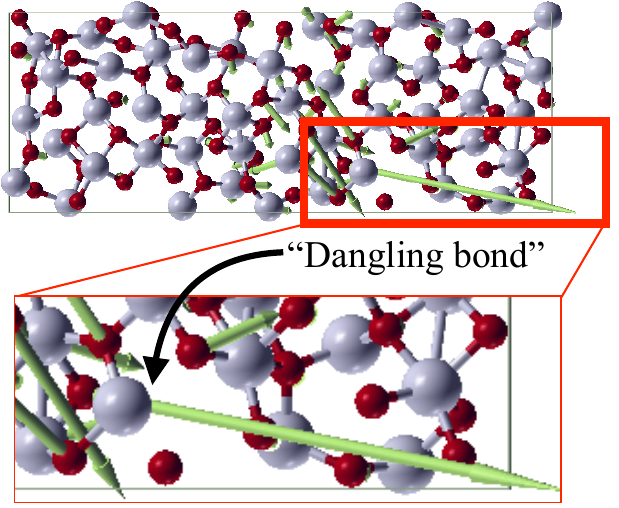}
    \caption{Enlarging the subset of atoms with maximal displacement vectors for lowest frequency phonon mode of amorphous Al$_{2}$O$_{3}$ sample, revealing that the largest displacement vector corresponds to an Al atom which exhibits increased isolation in real-space from the sample, allowing for a dangling bond-like feature. {\color{black} These structural features are absent in the crystalline phase, giving rise to the observed soft modes identified as TLS candidates.}}
    \label{fig:Zoom}
\end{figure}

\par 
Finally, the structure which is stable at $300K$ is the relaxed using density functional theory at zero temperature using a plane-wave cutoff of 50 $Ry$ and a $8 \times 4 \times 2$ grid of $\mathbf{k}$-points. We use such a fine mesh in relaxation to increase accuracy and avoid spurious negative frequency modes in the subsequent phonon computations. The atomic positions are relaxed until the system satisfies the constraint that forces are all less than $10^{-5}$ Ry/Bohr; this is referred to as the final structure. This process was used to generate five samples for which the structure files are available on GitHub. Examples of the final structures are shown in Fig. \eqref{fig:samples}.

\section{Low-frequency phonon modes}
We now examine the phonon spectra of each sample to search for evidence to support or reject the hypothesis that vibrons exist, localized in real space, which serve as TLSs coupling to the qubit. To compute the low-frequency phonon modes we utilize density functional perturbation theory (DFPT) within Quantum Espresso. We limit our computation of phonon modes to the $\Gamma$ location as the system is not an ordered solid. The lowest-energy phonon modes for each of the samples along with the phonon density of states upon averaging over each sample are shown in Fig. \eqref{fig:LowFModes}.
\par 
{\color{black}Examining the density of states we note a peak in the range $\approx 0.2-0.5$ THz. This peak corresponds directly to the vibron modes for which the displacement vectors are shown in \eqref{fig:LowFModes}(a)-(e).} It is important to address that this is not the expected frequency of $\sim 10$ GHz expected for a TLS which couples strongly to the qubit. However, this is not entirely surprising and is likely due in part to finite size effects. Low-frequency modes are long wavelength and may be artificially hardened by size constraints and artificial periodicity invoked by density functional theory. Future work is required to accommodate $>1000$ atoms, creating a unit cell sufficiently large to limit the effects of artificial periodicity. Machine learned interatomic potentials appear ideally positioned to tackle this future challenge\cite{chen2022universal,deng2023chgnet,batatia2022mace}.

\par 
{\color{black}As TLSs interact with qubits primarily through their electric dipole moment\cite{lisenfeld2015observation}, we compute the dipole moment of the identified vibron modes. Experimental studies have determined the dipole moment of TLSs in amorphous aluminum oxide, making it a suitable metric to determine the validity of the computational results.} The dipole moment of each mode is computed and displayed along with the respective phonon mode in Fig. \eqref{fig:LowFModes}. The dipole moment is computed as,
\begin{equation}
    p_{\alpha}=\sum_{\kappa}Z_{\kappa,\alpha}\Delta _{\kappa,\alpha}
\end{equation}
where $Z_{\kappa,\alpha}$ is the $\alpha$ component of the Born effective charge tensor for atom $\kappa$ computed via DFT and $\Delta _{\kappa,\alpha}$ is similarly the displacement along $\alpha$ for atom $\kappa$ for a given vibron. The expected dipole moment of a TLS in superconducting qubits has been estimated to be $\sim 0.54 e \AA$\cite{PhysRevApplied.17.034025}. This is in excellent agreement with the dipole moment computed for each of the vibrons, providing evidence towards their classification as dipole activated TLSs. 
\par 
{\color{black}In order to qualitatively understand the structural origin of these low-frequency modes we can zoom in on the region for which the displacement vectors are maximal. Doing so, we observe that these large displacement vectors occur for atoms which exhibit spatial isolation to a greater extent than their neighbors. To quantify the real-space localization of the low-frequency modes in real-space, we compute the inverse participation ratio as a function of frequency. The inverse participation ratio is computed as, 
\begin{equation}\label{eq:ipr}
    IPR(\alpha(\omega))= \frac{\sum_{i=1}^{120}|\mathbf{e}^{\alpha}_{i}|^4}{\left(\sum_{i=1}^{120}|\mathbf{e}^{\alpha}_{i}|^{2}\right)^{2}},
\end{equation}
where for phonon mode $\alpha$, $\mathbf{e}^{\alpha}_{i}$ is the normalized displacement vector at atom $i$. If a given phonon mode is localized to a single atom then this value approaches $1$, while a fully delocalized phonon mode yields $1/N$. We compute the IPR for each phonon mode and amorphous sample. Collecting all values of the IPR as a function of the frequency of the phonon mode, $\omega$, we can compute a density of states for the IPR as a function of frequency. The results are used to color code the phonon density of states  in Fig. \eqref{fig:LowFModes}(f). This color coding reveals the extent to which the low-frequency modes are localized compared to those at high-frequency. Due to this spatial isolation we can be consider these states to be akin to ``dangling bonds". }  The existence of dangling bonds within the bulk is a unique property of the amorphous systems as an ordered crystalline material will ensure that no atom is isolated in this manner. 
\par 
{\color{black} We have therefore successfully computed the low-frequency modes for amorphous samples of Al$_{2}$O$_{3}$ and found evidence that vibron modes, appearing as dangling bonds support a dipole moment in agreement with that experimentally measured. However, it is important to briefly discuss the manner in which such dangling bonds can give rise to a TLS. This is explored in Ref. \cite{varma2023theory}, where it is stated that the spatial isolation of a dangling bond enables a rotational degree of freedom, absent in a periodic crystal. As the environment surrounding the dangling bond is not uniform in the amorphous system, this rotational degree of freedom will produce an anharmonic potential well, providing correspondence between the TLS and dangling bond.} 

\begin{figure}
  \centering
    \includegraphics[width=8cm]{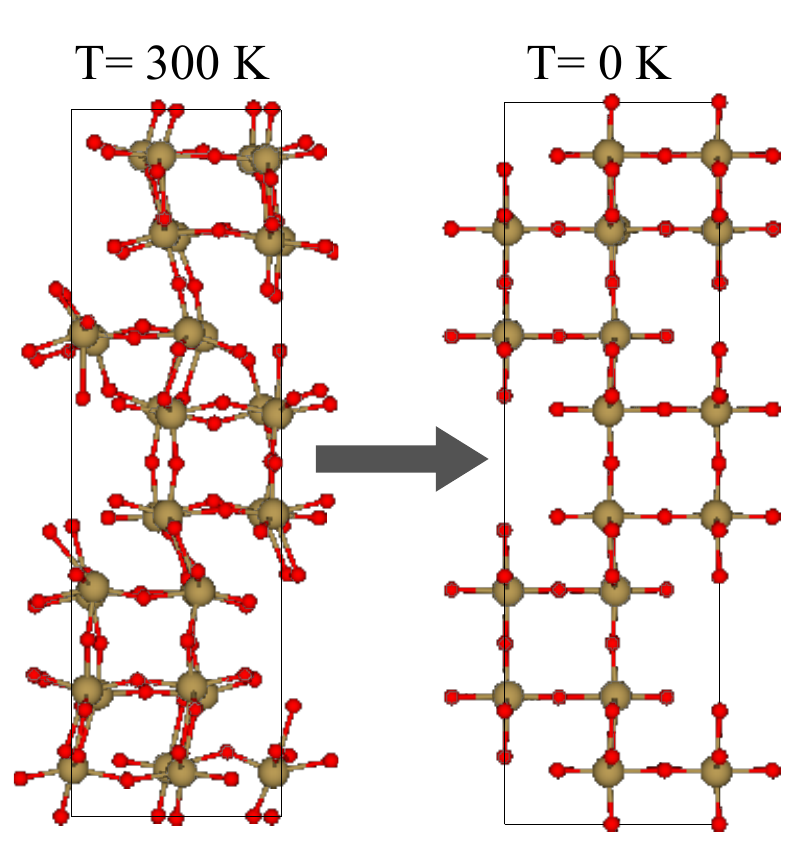}
    \caption{(Left) Structure of amorphous Ta$_{2}$O$_{5}$ at room temperature after undergoing a melting/cooling cycle and being equilibrated. (Right) Upon relaxation of the structure at 0K the system returns to the crystalline, periodic structure in contrast to amorphous Al$_{2}$O$_{3}$ which lacks long-range order after relaxation at 0K.}
    \label{fig:taox}
\end{figure}
\section{Mitigation strategies}
Identification of low-frequency modes, vibrons, as a source of TLSs in superconducting qubits provides an opportunity to determine mitigation strategies. Namely, we have established that these vibrons arise due to the amorphous nature of the system, allowing for the existence of dangling bond-like structures within the bulk of the sample. One route to remove these defects is to work with a compound admitting increased order, such as a quasi-crystal, rather than an amorphous solid. With increased order isolated defects can be minimized. 
\par 
It appears that amorphous oxides formed with heavier elements, such as TaO$_{x}$, are ideally positioned to reduce TLS density in modern superconducting qubits. The presence of heavier elements changes the energy landscape such that atomic configuration deviating from the convex hull can incur a greater energetic cost. It is expected that, when relaxed at zero temperature, amorphous TaO$_{x}$ has a greater likelihood of returning to a crystalline or quasi-crystalline form. To investigate this claim we repeat the process utilized for generation of amorphous Al$_{2}$O$_{3}$, applying it to a supercell of Ta$_{2}$O$_{5}$ containing 112 atoms. The same procedure is appropriate as the melting point of Ta$_{2}$O$_{5}$ is similar to that of Al$_{2}$O$_{3}$ at $2145K$\cite{PhysRevMaterials.3.055605}. However, due to the increased mass of Ta versus Al, when the system is equilibrated, either in the liquid phase or at any other point, we choose to double the number of time steps for which the system is equilibrated. When equilibrating in the liquid phase, we also impose an ion temperature of $3000K$ to aid in vanquishing any remnant of the crystal structure.  
\par 
Examining the final structures after equilibration at room temperature and zero temperature relaxation, shown in Fig. \eqref{fig:taox}, it is striking that the systems appears to have returned to the ordered crystalline phase. This is in stark contrast to the amorphous Al$_{2}$O$_{3}$ structures which do not resemble the ordered solid counterpart. While this computation suggests increased order in samples of amorphous Ta$_{2}$O$_{5}$ relative to amorphous Al$_{2}$O$_{3}$, future work is required to ensure this observation remains valid in the thermodynamic limit.
\par 
Despite the need for future work we remark that a separate study has examined the surfaces of amorphous TaO$_{x}$ structures, revealing a decreased density of TLSs when compared to the lighter amorphous NbO$_{x}$\cite{wang2024superconducting}. This is in agreement with our observations that the density of surface dangling bonds may be reduced in heavier oxides. Evidence of this is provided in the appendix where the surface density of states for amorphous Al$_{2}$O$_{3}$ and amorphous Ta$_{2}$O$_{5}$ are compared. 

\par
\section{Discussion}
In this work we have performed detailed first-principles computations to construct multiple samples of amorphous Al$_{2}$O$_{3}$. Computing the phonon modes we identify low-frequency modes exhibiting real-space localization and a dipole moment in agreement with that expected for TLSs in superconducting qubits which use amorphous Al$_{2}$O$_{3}$ as the Josephson junction barrier. {\color{black} Importantly, we find that the low-frequency modes, which we refer to as vibrons, take the form of dangling-bonds and are thus a unique feature of the non-crystalline system. This observation is in-line with prior works\cite{wang2019low,varma2023theory}, which have shown that a dangling bond may possesses a rotational degree of freedom not found in an ordered solid. This rotational degree of freedom gives rise to an anharmonic potential well, i.e. TLS. As the dangling bonds arise due to a lack of order, they can be mitigated by use of crystalline or quasi-crystalline materials for which such bulk defects are suppressed.} The computationally available energies we probe are in sub THz range. We do believe that similar excitations persist down to GHz range, relevant for device applications, and thus are  viable candidates for TLS dephasers in SC qubits. This observation also applies to the existence of TLSs on the surfaces.

We also compare the amorphous amorphous Al$_{2}$O$_{3}$ with the Ta$_{2}$O$_{3}$. We find that with similar protocol for sample preparation the 
Ta$_{2}$O$_{3}$ forms a regular crystalline structure and thus inherently is less disordered. We suggest that Ta$_{2}$O$_{3}$ as an optimal alternative. 

Future work is required to establish the precise quai-crystalline structure of Ta$_{2}$O$_{3}$ as synthesized in current superconducting qubits for comparison with first-principles analysis. In addition further samples and larger system sizes are required to further limit any finite size effects.

\par 
\acknowledgments{} 
We are grateful to D. Pappas, Y. Rosen, J. Mutus and X. Wang for useful discussions.  Work at NORDITA is supported by the Novo Nordisk Foundation and NordForsk. AB and JH were 
 also supported by European
Research Council under the European Union Seventh
Framework ERS-2018-SYG 810451 HERO and by 
Knut and Alice Wallenberg Foundation Grant No. KAW
2019.0068. The computations were enabled by resources provided by the National Academic Infrastructure for Supercomputing in Sweden (NAISS), partially funded by the Swedish Research Council through grant agreement no. 2022-06725. This work is supported by the Novo Nordisk Foundation, Grant number NNF22SA0081175, NNF Quantum Computing Programme

\bibliographystyle{apsrev4-1}
\nocite{apsrev41Control}
\bibliography{ref.bib}
\par 
\appendix
\section{Analysis of electronic density of states}
In the main body we focus primarily on the identification of soft modes in the phonon spectra as potential two-level systems (TLSs), hastening decoherence in modern superconducting qubits. Computation of the phonon modes was done for bulk samples of amorphous Al$_{2}$O$_{3}$ (amorphous Al$_{2}$O$_{3}$). This is necessary due to the immense computational expense associated with the generation of large amorphous Al$_{2}$O$_{3}$ samples within first-principles, and subsequent computation of the phonon modes. Despite computational considerations limiting analysis of phonon soft modes to the bulk, it is important to consider the surfaces of amorphous Al$_{2}$O$_{3}$. Surfaces can support a high-density of dangling bonds manifesting as surface roughness. The analysis presented in the main body validates the hypothesis that dangling-bonds and similar structurally isolated features in an amorphous system can give rise to localized soft-modes which serve as two-level systems.
\begin{figure}
    \centering
    \includegraphics[width=8cm]{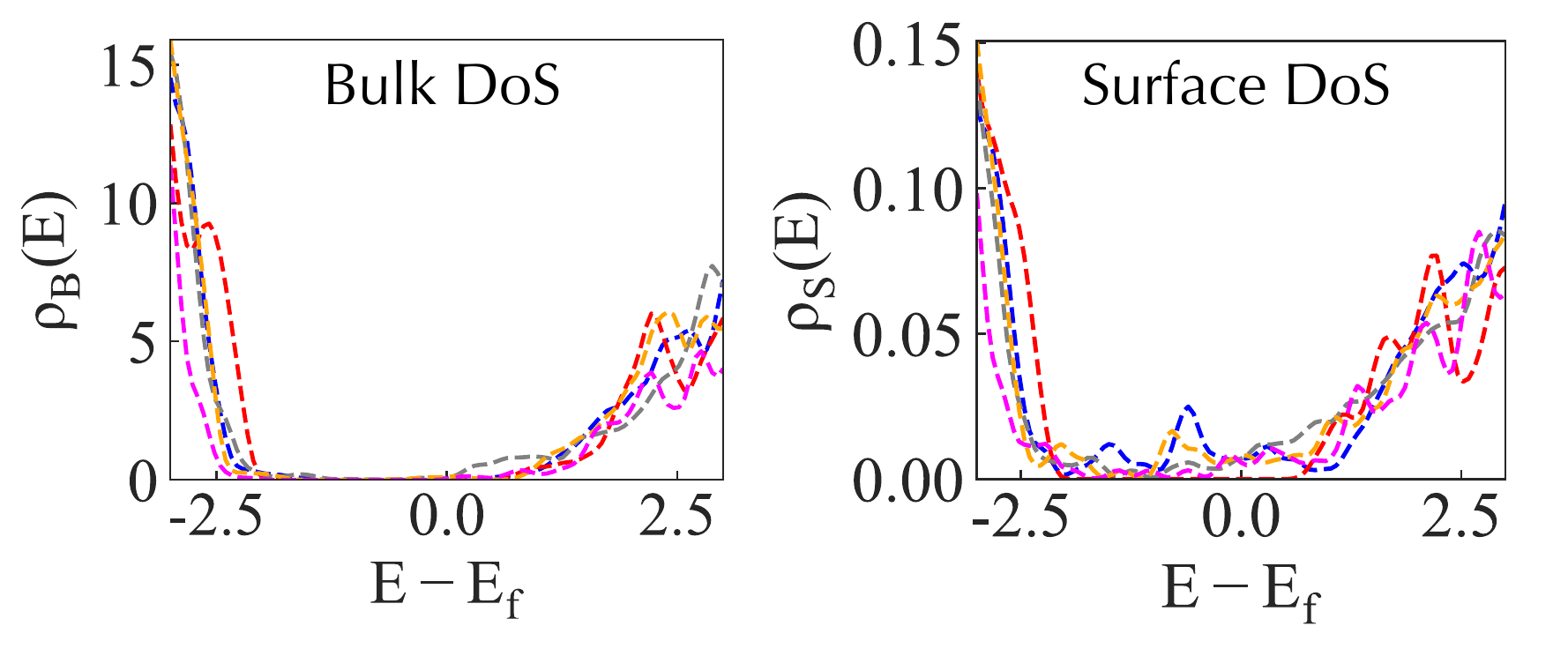}
    \caption{(left) Bulk density of states for five distinct samples of amorphous Al$_{2}$O$_{3}$. Each displays a robust gap in the electronic spectra at the Fermi level. (right) Density of states for five distinct samples of amorphous Al$_{2}$O$_{3}$ upon applying open boundary conditions along both the $x$ and $y$ directions, revealing emergent, surface bound electronic states within the mid-gap.}
    \label{fig:DoSAlOx}
\end{figure}
\begin{figure*}
    \centering
    \includegraphics[width=16cm]{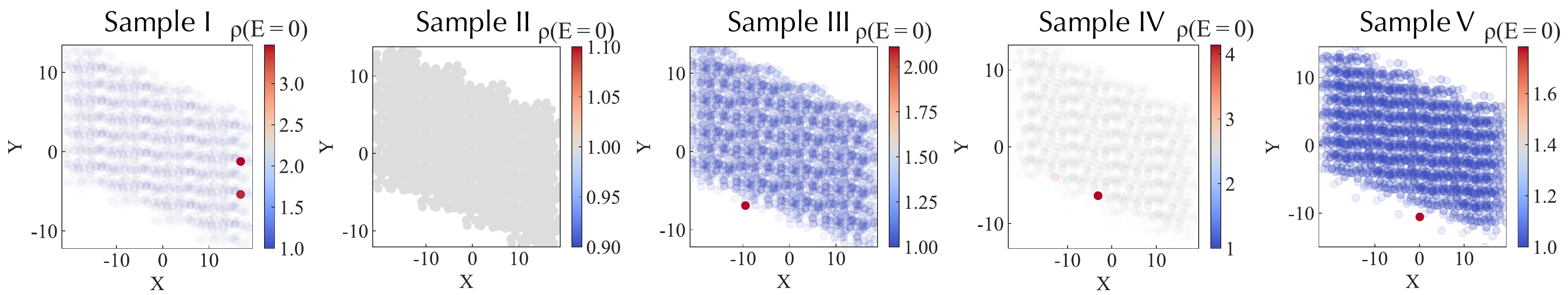}
    \caption{Spatially resolved local density of states at the Fermi level for five distinct samples of amorphous Al$_{2}$O$_{3}$ upon application of open-boundary conditions along the $x$ and $y$ directions. Red points mark locations of greatest spectral density, they are located on the sample and spatially isolated in each case, revealing that the low-energy density of states seen in the right panel of Fig. \eqref{fig:DoSAlOx} are localized surface modes.}
    \label{fig:DoSAlOx_RS}
\end{figure*}

\begin{figure}
    \centering
    \includegraphics[width=8cm]{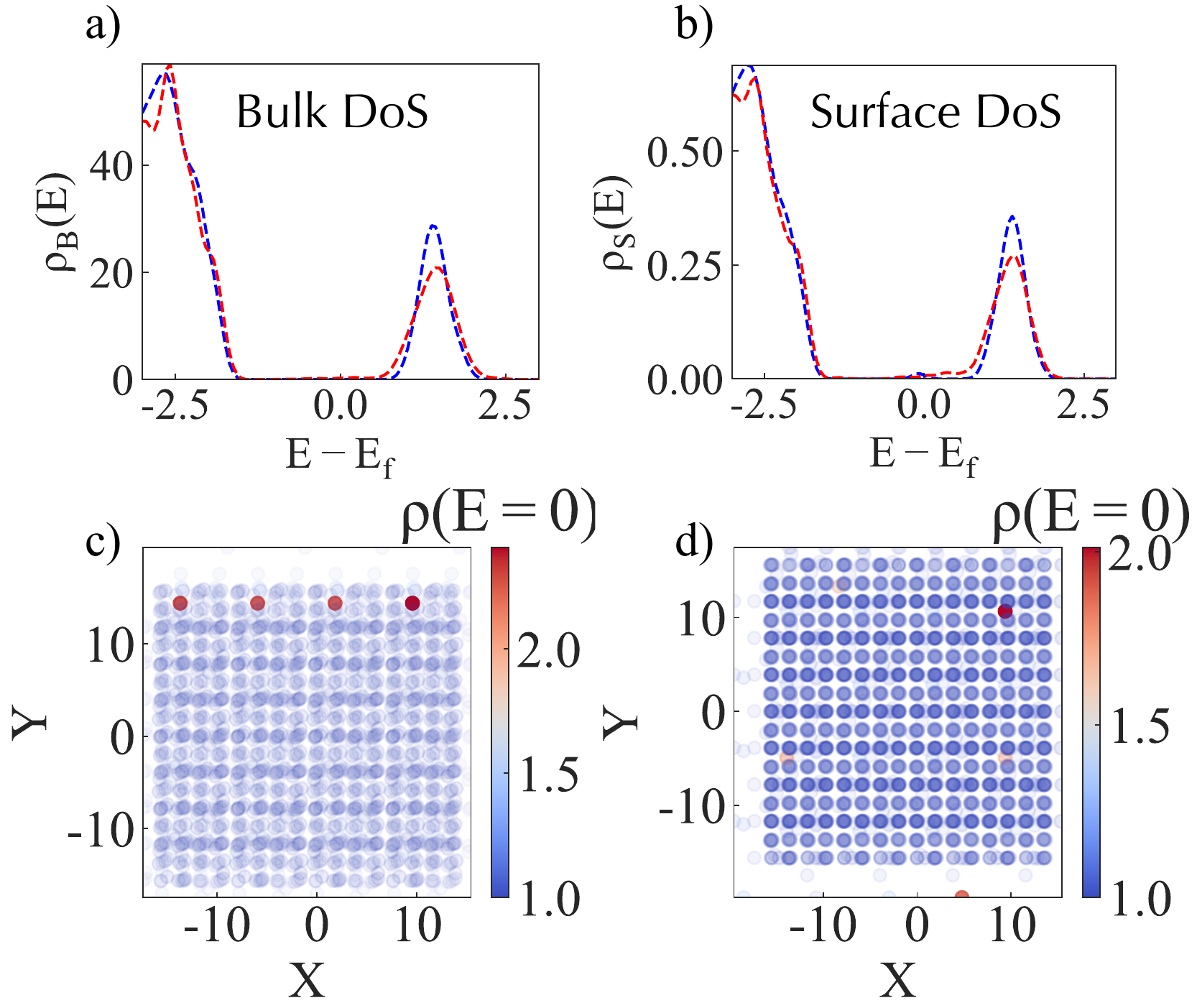}
    \caption{(a) Bulk density of states for two distinct samples of amorphous Ta$_{2}$O$_{5}$. Each displays a robust gap in the electronic spectra at the Fermi level. (b) Density of states for the same samples of amorphous Ta$_{2}$O$_{5}$ when applying open boundary conditions along both the $x$ and $y$ directions. While the gap is softened, the density of surface bound mid-gap electronic states is reduced compared to amorphous Al$_{2}$O$_{3}$. (c)-(d) The spatially resolved density of states at the Fermi energy in the $x$-$y$ plane for the two samples of  amorphous Ta$_{2}$O$_{5}$, demonstrating that any mid-gap states remain surface bound.}
    \label{fig:DoSTaOx}
\end{figure}
\par 
{\color{black}Following analysis laid out in the main body, dangling-bind like features are a key structural indicator of localized soft modes which are TLS candidates. Therefore, dangling bonds on the surface of amorphous Al$_{2}$O$_{3}$ are likely candidates for soft phonon modes, i.e. TLSs. Additionally, dangling bonds on the surface can support low-energy electronic bound states. This is well studied in prior works on structural disorder\cite{schofield2013quantum,wang2024superconducting}.} As such, probing the electronic density of states on the surface can be used to determine surface roughness and the density of TLSs. In order to model the surface of amorphous Al$_{2}$O$_{3}$ and compute the density of states, we construct real-space tight-binding models for each of the amorphous Al$_{2}$O$_{3}$ samples generated in the main body using the Wannier90 software package\cite{Pizzi2020}. The bulk electronic states of the tight-binding model precisely matches that computed from first-principles in each case and is shown for the five amorphous samples in Fig. \eqref{fig:DoSAlOx_RS}. In each case we note the presence of a clean, insulating band-gap in line with expectations from experimental studies.

\par 
Utilizing the tight-binding models, we then consider the electronic density of states upon imposing open boundary conditions along the $x$ and $y$ directions and recompute the electronic density of states. The results shown in the left panel of Fig. \eqref{fig:DoSAlOx} detail the emergence of low-energy mid-gap electronic state, which must be localized on the surface when comparing to the results shown in the right panel of Fig. \eqref{fig:DoSAlOx}. To investigate this further, we compute the spatially resolved local density of states at zero energy with the results shown in Fig. \eqref{fig:DoSAlOx_RS}. These figures demonstrate that the resulting low-energy electronic states on the surface are localized, not extended, in correspondence with the presence of surface roughness and dangling bonds. 

\par 
For completeness, we compare the low-energy electronic density of states on the surface with that computed for amorphous Ta$_{2}$O$_{5}$. The bulk electronic density of states for two samples of amorphous Ta$_{2}$O$_{5}$ are shown in Fig. \eqref{fig:DoSTaOx} along with the electronic density of states upon imposing open boundary conditions along the $x$ and $y$ directions. In agreement with the observation that amorphous Ta$_{2}$O$_{5}$ supports a more ordered structure, reducing surface roughness and the density of TLSs, the magnitude of mid-gap electronic density of states is reduced compared with amorphous Al$_{2}$O$_{3}$. This is in alignment with the conclusions reached in Ref. \cite{wang2024superconducting}, that the surface of amorphous Ta$_{2}$O$_{5}$ will support a reduced density of TLSs.

\end{document}